\documentclass[prl,twocolumn]{revtex4}

\usepackage{graphicx}
\usepackage{amsmath}
\usepackage{ifthen}
\usepackage{latexsym}

\newcommand{\vekt}[1]{\boldsymbol{#1}}

\begin{document}

\title{High-fidelity entanglement swapping with fully independent sources}

\author{Rainer Kaltenbaek}
\thanks{Present address: Institute for Quantum Computing and Department of Physics and
Astronomy, University of Waterloo, Waterloo, Canada, N2L 3G1}
\affiliation{Faculty of Physics, University of Vienna, Boltzmanngasse 5, A-1090 Vienna, Austria}
\author{Robert Prevedel}
\affiliation{Faculty of Physics, University of Vienna, Boltzmanngasse 5, 
             A-1090 Vienna, Austria}
\author{Markus Aspelmeyer}
\affiliation{Institute for Quantum Optics and Quantum Information (IQOQI), 
             Austrian Academy of Sciences, Boltzmanngasse 3, A-1090 Vienna, 
	     Austria}
\author{Anton Zeilinger}
\email{anton.zeilinger@univie.ac.at}
\affiliation{Faculty of Physics, University of Vienna, Boltzmanngasse 5, 
             A-1090 Vienna, Austria}
\affiliation{Institute for Quantum Optics and Quantum Information (IQOQI), 
             Austrian Academy of Sciences, Boltzmanngasse 3, A-1090 Vienna, 
	     Austria}

\pagestyle{empty}

\begin{abstract}
Entanglement swapping allows to establish entanglement between 
independent particles that never interacted nor share any common past. This 
feature makes it an integral constituent of quantum repeaters.
Here, we demonstrate entanglement swapping with time-synchronized independent
sources with a fidelity high enough to violate a Clauser-Horne-Shimony-Holt 
inequality by more than four standard deviations. The fact that 
both entangled pairs are created by fully independent, only electronically 
connected sources ensures that this technique is suitable for
future long-distance quantum communication experiments as well as for 
novel tests on the foundations of quantum physics.
\end{abstract}

\maketitle
To take full advantage of the powerful features of quantum information 
processing schemes~\cite{Bouwmeester2000a}, large scale quantum 
networks will have to be realized. These will inevitably involve the reliable 
distribution of entanglement~\cite{Schroedinger1935a} between distant, 
independent nodes. Photons are well suited to cover these distances as they 
do not tend to interact with the environment and can easily be 
transmitted via fibers or optical free-space links.
Today, these methods are limited to distances on the order of
a hundred kilometers~\cite{Marcikic2005a,Huebel2007a,Ursin2007a} due to 
photon loss and detector noise~\cite{Waks2002a}. 
Quantum repeaters~\cite{Briegel1998a,Duer1999a,Duan2001a,
Sangouard2008a}, which repeatedly swap and distill entanglement
along a chain of distant sources, are expected to overcome this limitation. 

The three principal ingredients of quantum repeaters are quantum 
memories~\cite{Duan2001a,Sangouard2008a}, entanglement 
distillation~\cite{Bennett1996a,Deutsch1996a}, and entanglement 
swapping~\cite{Zukowski1993a}. 
While a real-world implementation of a quantum repeater is still far down the 
road, gradual progress has been achieved in realizing its 
constituents~\cite{Pan1998a,Jennewein2002a,Pan2003a,Matsukevich2004a,
Jennewein2004a,Walther2005a,Yuan2008a,Choi2008a,Matsukevich2008a}. 
In particular any future realization of a 
quantum repeater will involve entanglement swapping with pairs emitted from 
distant, independent sources. 

Recently, the independence of the sources used for entanglement 
swapping was shown to be not only of practical interest. 
In~\cite{Greenberger2008b} it is argued that independence in such an experiment
allows to place even tighter restrictions on local hidden variable 
theories~\cite{Bell1964a} 
than in experiments on pairs emitted directly by one source. This
allows to circumvent the loophole of inefficient 
detectors. Therefore, the use of independent sources for 
entanglement swapping could help to close all loop holes, which up to now had 
to be closed in separate experiments~\cite{Aspect1982b,Weihs1998a,Rowe2001a,
Matsukevich2008a}, in one ultimate Bell-type experiment.

Entanglement swapping has been demonstrated with 
photons~\cite{Pan1998a,Jennewein2002a,Jennewein2004a,Yang2006a} and 
recently with atomic ensembles~\cite{Yuan2008a} and 
ions~\cite{Matsukevich2008a}. In all of these experiments a crucial 
requirement for use in quantum repeaters remains unfulfilled: the sources 
must be separable by large distances. 
In~\cite{Pan1998a,Jennewein2002a,Jennewein2004a,Yuan2008a,Matsukevich2008a} 
the entangled pairs were created by interaction of one optical pump with 
one or two nonlinear media, while in~\cite{Yang2006a} the two laser beams 
pumping two separate nonlinear media were not optically independent. To 
separate the sources in any of these implementations
would require to distribute intense pump beams over large distances. That
goal is practically unfeasible given dispersion, high loss at the pump 
wavelengths and path length fluctuations. Moreover, because of either 
the common origin or the optical interaction of the pump beams involved, 
in none of these experiments the sources meet the criteria for independence 
as needed for a real-life quantum repeater or for an experiment as 
presented in~\cite{Greenberger2008b}.

Previous experiments~\cite{Kaltenbaek2006a,Halder2007a} aimed to show that
entanglement swapping with fully independent sources is in principle feasible.
However, these experiments failed to show the genuine none-classical
correlations of the swapped entangled states. In particular, while correlations
were observed in~\cite{Halder2007a}, they where not strong
enough to violate a Bell-type inequality~\cite{Bell1964a}, which 
would indicate that the swapped entanglement was sufficient for direct 
further use in, for example, quantum-state teleportation~\cite{Bennett1993a} 
or entanglement-based cryptography~\cite{Ekert1991a,Bennett1992b}
without the need for distillation.

Here, we fill this experimental gap and demonstrate high-fidelity entanglement 
swapping between entangled photon pairs emitted from time-synchronized 
independent sources. The resulting correlations between particles that do 
not share any common past are strong enough to violate
a Clauser-Horne-Shimony-Holt (CHSH) inequality~\cite{Clauser1969a}.
Our configuration is a prototype solution for use in future quantum
repeaters. It is readily adaptable for use over large distances, and 
it implements a BSM with the maximum achievable efficiency for a 
linear-optics implementation.

\begin{figure}
 \begin{center}
 \includegraphics[width=1.0\linewidth]{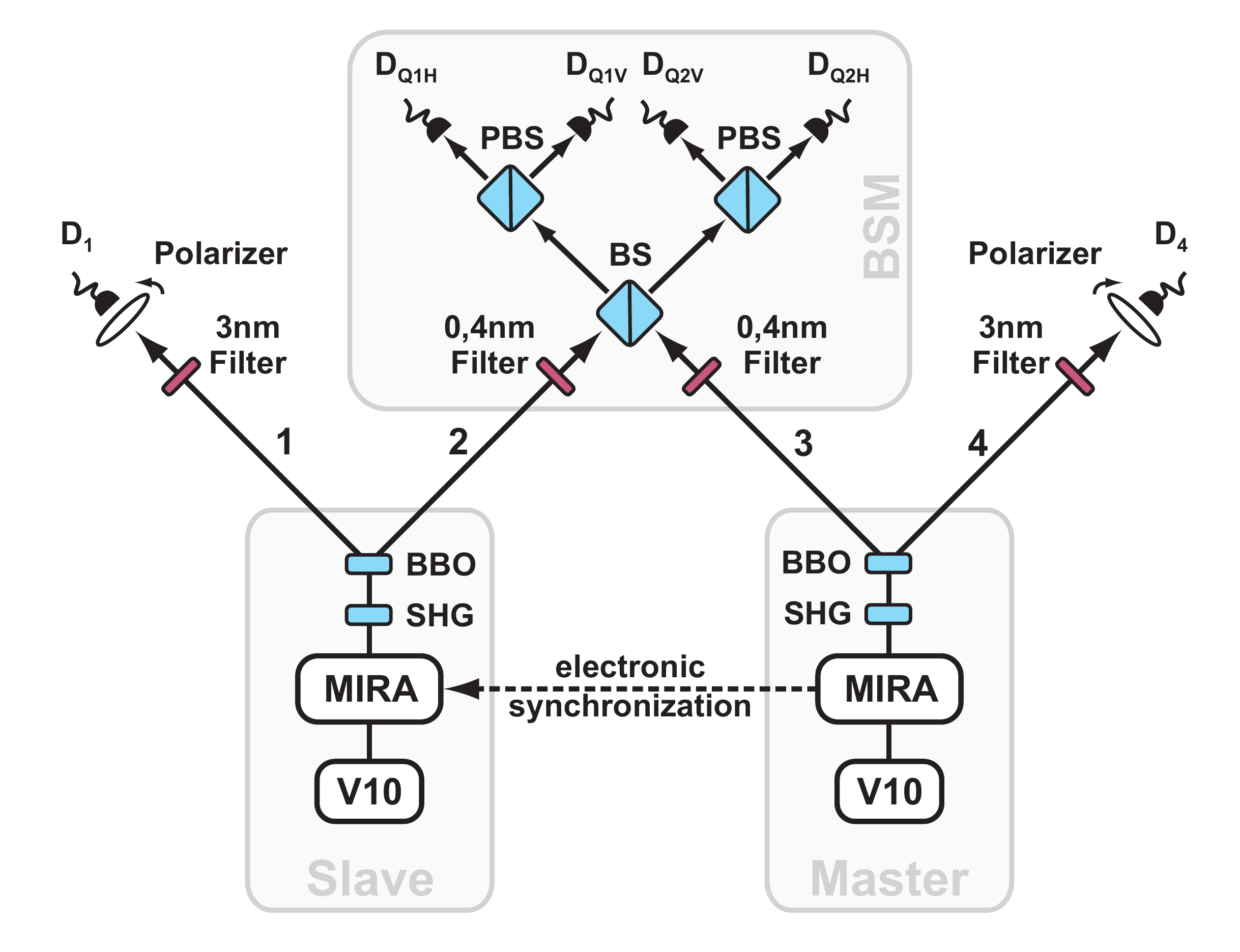}
 \caption{Schematic of the setup for entanglement swapping. The two 
 femtosecond lasers, which are separated by 
 $\sim 4\,\mbox{m}$, are time-synchronized by a Coherent Synchrolock
 (for details see ref.~\cite{Kaltenbaek2006a}).
 Each of the two infrared beams is frequency doubled to $394.25\,\mbox{nm}$
 by second-harmonic generation (SHG), the master's beam by a 
 Lithium-Triborate crystal and the slave's beam by a $\beta$-Barium-Borate 
 (BBO)
 crystal. The resulting beams each pump a BBO crystal to generate entangled 
 photons via SPDC. Interference 
 filters of $0.4\,\mbox{nm}$ and  $3\,\mbox{nm}$ FWHM bandwidth guarantee
 high HOM interference visibility for the BSM, which is realized by a 
 combination of two polarizing  beam splitters (PBS) and a fiber beam 
 splitter (BS)\label{fig::setup}.}
 \end{center}
\end{figure}

Each source in our experiment emits pairs of polarization entangled 
photons along spatial directions $1$ \& $2$ and $3$ \& $4$, respectively (see 
fig.~\ref{fig::setup}). 
We chose the singlet state $\psi^-$, which is one of the four maximally 
entangled Bell states:
\begin{equation}
\begin{array}{l}
\vert \psi^{\pm} \rangle = \frac{1}{\sqrt{2}} \vert HV\rangle \pm \vert VH\rangle \\
\vert \phi^{\pm} \rangle = \frac{1}{\sqrt{2}} \vert HH\rangle \pm \vert VV\rangle.
\end{array}
\end{equation}
A successful entanglement swapping procedure will result in photons $1$ and $4$
being entangled, although they never interacted with each 
other~\cite{Zukowski1993a,Jennewein2002a}. This is done by performing a 
Bell-state 
measurement on particles $2$ and 
$3$, i.e.~by projecting them on one of 
the four Bell states. Consequently, photons $1$ and $4$ will be projected onto
the Bell state corresponding to the BSM outcome.
Because of the independence of the sources, either the emission or the 
detection of the 
individual photons has to be time-synchronized better than the coherence 
times of the photons~\cite{Zukowski1995a}. One possible method 
to fulfill this requirement is to actively time-synchronize pulsed 
sources~\cite{Kaltenbaek2006a}. In contrast to techniques used in 
earlier experiments~\cite{Pan1998a,Jennewein2002a,Jennewein2004a,Yang2006a,
Yuan2008a,Matsukevich2008a} this allows to separate the sources, in principle, 
by arbitrary distances.

Each of the two separate spontaneous parametric down-conversion (SPDC) setups
is pumped by a frequency-doubled beam ($\lambda=394.25\,\mbox{nm}$) 
generated by two separate, pulsed femtosecond lasers (Coherent MIRA, 
operating at $788.5\,\mbox{nm}$), each of which is in turn pumped by its 
own solid-state laser (Coherent Verdi V10). The timing of the pump pulses is
synchronized purely electronically to an accuracy of $260\pm30\,\mbox{fs}$ 
using the method described in~\cite{Kaltenbaek2006a}. No optical interaction 
between the two sources is necessary to sustain time-synchronization. 
The synchronization mechanism consists of two phase-locked loops (PLL), one
for coarse and one for tight synchronization. The inputs of the PLLs are
the signals of two fast photodiodes ($2\,\mbox{GHz}$ bandwidth), each 
monitoring the output of one of the lasers. For details 
see~\cite{Kaltenbaek2006a}. The feedback signals of the two PLLs that adjust 
the cavity length of the slave laser have a bandwidth 
$\Delta\nu \le 10\,\mbox{kHz}$. This restricts the maximum distance of the
two sources to $c/\Delta\nu \approx 30\,\mbox{km}$. Here, we assume that the 
PLLs are positioned at the slave laser, and that only the master laser's 
photodiode signal has to be transmitted over the separating distance, e.g.~via
a radio-frequency fiber-optic link. In practice, the maximum distance is 
additionally limited by path-length fluctuations and noise in the 
transmission of the diode signal.

We implement the Bell-state measurement by Hong-Ou-Mandel (HOM) interference 
at a 50:50 beam splitter~\cite{Hong1987a} and subsequent 
polarizing beam splitters~\cite{Braunstein1995a,Jennewein2004a} (see 
fig.~\ref{fig::setup}). It allows to identify two out of four Bell 
states, as was first demonstrated in the experimental realization of 
dense coding~\cite{Mattle1996a}. This is the 
optimum efficiency possible with linear optics~\cite{Calsamiglia2001a}. 
A two-fold coincidence detection
event between either $D_{Q1H}$ and $D_{Q2V}$ or $D_{Q1V}$ and $D_{Q2H}$
indicates a projection on $\psi^-$. On the other hand, a coincidence detection
event between either $D_{Q1H}$ and $D_{Q1V}$ or $D_{Q2H}$ and $D_{Q2V}$
indicates a projection on $\psi^+$. These events have to occur in coincidence 
with clicks in the detectors $D_1$ and $D_4$ in modes $1$ and $4$. 
All measurement results are, therefore, four-fold coincidence detection 
events, where the coincidence window has to be shorter than the delay between 
two successive pulses ($\sim 13\,\mbox{ns}$). 

\begin{figure}
 \begin{center}
 \includegraphics[width=1.0\linewidth]{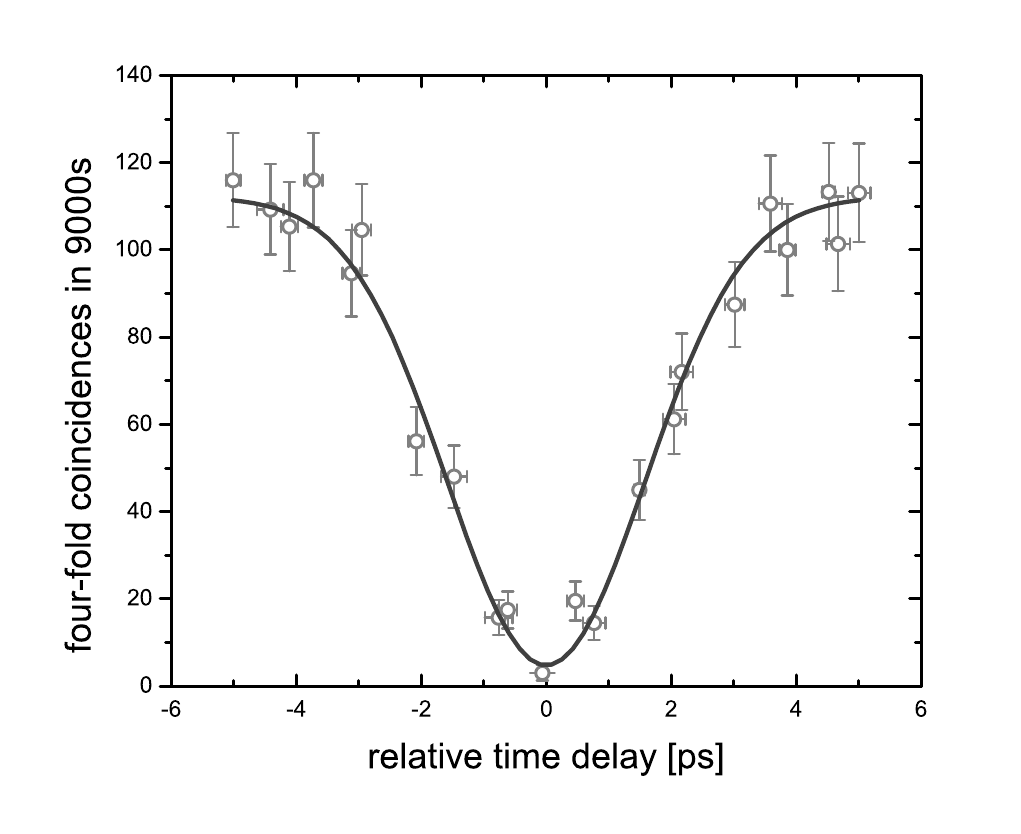}
 \caption{Hong-Ou-Mandel (HOM) interference scan. The 
        four-fold coincidences are measured between the detectors 
	$D_1$, $D_4$ and two detectors placed directly behind the fiber 
	beam splitter in the BSM (see Figure~\ref{fig::setup}). These are 
	plotted over the relative time delay of the interfering photons, which
	is determined via cross correlation of the laser pulses. 
	Background has not been
	subtracted. Error bars indicate
	s.d. The solid line is a Gaussian fit to the data ($\chi^2\sim 0.85$) 
	with a visibility of $96\pm 1\,\%$.\label{fig::dip}.}
 \end{center}
\end{figure}

Compared to~\cite{Kaltenbaek2006a} and~\cite{Halder2007a} we observed a 
significantly higher HOM-interference visibility of $(I_{max}-I_{min})/I_{max} = 0.96\pm 0.01$ (see fig.~\ref{fig::dip}). This was achieved by using
narrower bandwidth filters ($0.4\,\mbox{nm}$ FWHM). Using solid-state 
pump lasers in both of the sources, and stabilizing the laboratory 
temperature allowed us to dispense with the compensation for long-time drifts 
that was necessary in \cite{Kaltenbaek2006a}.
The considerably higher interference visibility observed was 
necessary in order to achieve the violation of a 
CHSH inequality~\cite{Clauser1969a} with the swapped entangled pairs.

We confirm successful entanglement swapping by testing the entanglement
of the previously uncorrelated photons $1$ and $4$. Violation of a CHSH 
inequality is not only of fundamental interest because it rules out 
local-hidden variable theories. It also proves that the swapped states are 
strongly entangled and, as a result, distillable~\cite{Horodecki1997a}. 

The specific state of photons $1$ and $4$ after entanglement swapping
depends on the result of the BSM, which can either be $\psi^+$ or $\psi^-$. 
The relevant CHSH inequalities for these cases are
\begin{eqnarray}
S_{\psi^{\mp}} & = & \left| \pm E(\vekt{a}_1,\vekt{b}_1) 
\mp E(\vekt{a}_1,\vekt{b}_2)\right. \nonumber\\
&  & \left.\;+\;E(\vekt{a}_2,\vekt{b}_1) + E(\vekt{a}_2,\vekt{b}_2)\right| \le
2\label{equ::CHSH}.
\end{eqnarray}
$E(\vekt{a}_i,\vekt{b}_i)$ denotes the expectation 
value of the product of the outcomes of coincidence measurements along 
$\vekt{a}_i$ and $\vekt{b}_i$ of the polarizations of the two photons. 
Each polarization measurement has two possible outcomes, which are assigned 
the values $+1$ and $-1$. In the experiment the vectors $\vekt{a}_i$ and 
$\vekt{b}_i$ 
correspond to the well known angles for the maximal violation of the CHSH 
inequality, $0^\circ$ and $45^\circ$ for photon $1$ and $22.5^\circ$ and 
$67.5^\circ$ for photon $4$, respectively.

\begin{table}
 \begin{center}
  \begin{tabular}{|c|c|c|c|c|}
  \hline &
  $E\left(0^\circ,22.5^\circ\right)$ &
  $E\left(0^\circ,67.5^\circ\right)$ &
  $E\left(45^\circ,22.5^\circ\right)$ &
  $E\left(45^\circ,67.5^\circ\right)$ \\
  \hline
  $\psi^-$ & $-0.53\pm 0.05$ & $0.65\pm 0.04$ & $-0.68\pm 0.04$ & 
             $-0.54\pm0.05$ \\
  \hline
  $\psi^+$ & $-0.44\pm 0.05$ & $0.69\pm 0.04$ & $0.69\pm 0.04$ & 
             $0.56\pm0.05$ \\
  \hline
  \end{tabular}
\caption{Expectation values for the CHSH inequality depending on
the outcome, $\psi^-$ or $\psi^+$, of the BSM. Each of these values is 
calculated from four
measurements of four-photon coincidences integrated over 
$15000\,\mbox{s}\sim 4\,\mbox{h}$.\label{tab::expvalues}}
\end{center}
\end{table}

The results for the expectation values for the various measurement settings
are given in table~\ref{tab::expvalues}. Applying them to 
equation~\ref{equ::CHSH} yields
\begin{equation}
\begin{array}{l}
S_{\psi^-} = 2.40 \pm 0.09 \\
S_{\psi^+} = 2.38 \pm 0.09,
\end{array}\label{equ::violation}
\end{equation}
where we have normalized the four-fold coincidence rates on the product of
two-fold coincidences to take into account degrading alignment over the long
scan times. Even without normalizing the four-folds we achieve a clear 
violation of the CHSH inequality with $S_{\psi^-} = 2.37\pm 0.09$ and 
$S_{\psi^+} = 2.38\pm 0.09$. 

In all cases the CHSH inequality is violated by more than four
standard deviations. This shows that the independent photons in modes $1$ and 
$4$ clearly are entangled and can asymptotically be distilled into the
maximally entangled singlet state~\cite{Horodecki1997a}, satisfying the 
central requirement for
quantum repeaters. Because every (two qubit) entangled state can be 
distilled but does not necessarily violate a Bell 
inequality~\cite{Werner1989a},
the violation of the CHSH inequality is a stronger criterion than required for 
distillability. It implies that, even without distillation, the entanglement of
the states obtained in our experiment is high enough for immediate further 
use in quantum 
communication protocols like quantum teleportation~\cite{Bennett1993a} or 
quantum cryptography~\cite{Ekert1991a,Bennett1992b}.

We have demonstrated high-fidelity entanglement swapping with 
time-synchronized independent sources. The swapped entanglement clearly 
violates a Bell-type inequality. These strong non-classical correlations 
between particles that do not share any common past are not only crucial
for future quantum repeaters. They might also enable novel tests of quantum 
mechanics~\cite{Yurke1992b,Greenberger2008a,Greenberger2008b}. Because the 
time-synchronization depends only on electronic signals, our sources 
can, in principle, be separated by very large distances. 
A practical way to do so would be to synchronize each laser locally to a 
reference oscillator, e.g.~an atomic clock. These reference oscillators can 
be synchronized remotely to sub-picosecond accuracy~\cite{Lutes1987a,
Celano2002a,Hudson2006a}, which would be sufficient for our purposes. Recent 
results~\cite{Kim2008a} show that even better remote timing
stability is achievable with state-of-the-art technology.
Future efforts should be directed at increasing the spectral
brightness of the SPDC sources used, e.g.~by employing pump lasers 
with narrower bandwidths and by using state-of-the-art SPDC 
sources~\cite{Fedrizzi2007a}. Only recently it was shown that SPDC bandwidths
narrow enough for quantum memories are feasible~\cite{Bao2008a}. As an 
added benefit, narrower bandwidths loosen
the constraints on laser and path-length synchronization to observe 
high-visibility HOM interference. Finally, a very important
step will be to efficiently produce and detect entangled photons at telecom 
wavelengths to cover even larger distances.
 
We are grateful to M.~Arndt, B.~Blauensteiner, H.~R. B\"ohm, Th.~Jennewein 
and M.~\. Zukowski for discussion and experimental advice. We acknowledge 
support from the Austrian Science Fund FWF, the City of Vienna, IARPA
(U.S.~Army Research Office) and the European Commission under projects 
SECOQC, EMALI and QAP. 

\bibliographystyle{apsrev}
\bibliography{swapping}

\begin{thebibliography}{49}
\expandafter\ifx\csname natexlab\endcsname\relax\def\natexlab#1{#1}\fi
\expandafter\ifx\csname bibnamefont\endcsname\relax
  \def\bibnamefont#1{#1}\fi
\expandafter\ifx\csname bibfnamefont\endcsname\relax
  \def\bibfnamefont#1{#1}\fi
\expandafter\ifx\csname citenamefont\endcsname\relax
  \def\citenamefont#1{#1}\fi
\expandafter\ifx\csname url\endcsname\relax
  \def\url#1{\texttt{#1}}\fi
\expandafter\ifx\csname urlprefix\endcsname\relax\def\urlprefix{URL }\fi
\providecommand{\bibinfo}[2]{#2}
\providecommand{\eprint}[2][]{\url{#2}}

\bibitem[{\citenamefont{Bouwmeester et~al.}(2000)\citenamefont{Bouwmeester,
  Ekert, and Zeilinger}}]{Bouwmeester2000a}
\bibinfo{author}{\bibfnamefont{D.}~\bibnamefont{Bouwmeester}},
  \bibinfo{author}{\bibfnamefont{A.}~\bibnamefont{Ekert}}, \bibnamefont{and}
  \bibinfo{author}{\bibfnamefont{A.}~\bibnamefont{Zeilinger}},
  \emph{\bibinfo{title}{The Physics\\ of Quantum Information}}
  (\bibinfo{publisher}{Springer}, \bibinfo{address}{Berlin},
  \bibinfo{year}{2000}).

\bibitem[{\citenamefont{Schr{\" o}dinger}(1935)}]{Schroedinger1935a}
\bibinfo{author}{\bibfnamefont{E.}~\bibnamefont{Schr{\" o}dinger}},
  \bibinfo{journal}{{D}ie {N}aturwissenschaften} \textbf{\bibinfo{volume}{48}},
  \bibinfo{pages}{807} (\bibinfo{year}{1935}).

\bibitem[{\citenamefont{Marcikic et~al.}(2004)\citenamefont{Marcikic,
  de~Riedmatten, Tittel, Zbinden, Legr\'e, and Gisin}}]{Marcikic2004a}
\bibinfo{author}{\bibfnamefont{I.}~\bibnamefont{Marcikic}},
  \bibinfo{author}{\bibfnamefont{H.}~\bibnamefont{de~Riedmatten}},
  \bibinfo{author}{\bibfnamefont{W.}~\bibnamefont{Tittel}},
  \bibinfo{author}{\bibfnamefont{H.}~\bibnamefont{Zbinden}},
  \bibinfo{author}{\bibfnamefont{M.}~\bibnamefont{Legr\'e}}, \bibnamefont{and}
  \bibinfo{author}{\bibfnamefont{N.}~\bibnamefont{Gisin}},
  \bibinfo{journal}{Phys. Rev. Lett.} \textbf{\bibinfo{volume}{93}},
  \bibinfo{pages}{180502} (\bibinfo{year}{2004}).

\bibitem[{\citenamefont{H{\"u}bel et~al.}(2007)\citenamefont{H{\"u}bel, Vanner,
  Lederer, Blauensteiner, Lor{\" u}nser, Poppe, and Zeilinger}}]{Huebel2007a}
\bibinfo{author}{\bibfnamefont{H.}~\bibnamefont{H{\"u}bel}},
  \bibinfo{author}{\bibfnamefont{M.~R.} \bibnamefont{Vanner}},
  \bibinfo{author}{\bibfnamefont{T.}~\bibnamefont{Lederer}},
  \bibinfo{author}{\bibfnamefont{B.}~\bibnamefont{Blauensteiner}},
  \bibinfo{author}{\bibfnamefont{T.}~\bibnamefont{Lor{\" u}nser}},
  \bibinfo{author}{\bibfnamefont{A.}~\bibnamefont{Poppe}}, \bibnamefont{and}
  \bibinfo{author}{\bibfnamefont{A.}~\bibnamefont{Zeilinger}},
  \bibinfo{journal}{Optics Express} \textbf{\bibinfo{volume}{15}},
  \bibinfo{pages}{7853} (\bibinfo{year}{2007}).

\bibitem[{\citenamefont{Ursin et~al.}(2007)\citenamefont{Ursin, Tiefenbacher,
  Schmitt-Manderbach, Weier, Scheidl, Lindenthal, Blauen\-steiner, Jennewein,
  Perdigues, Trojek et~al.}}]{Ursin2007a}
\bibinfo{author}{\bibfnamefont{R.}~\bibnamefont{Ursin}},
  \bibinfo{author}{\bibfnamefont{F.}~\bibnamefont{Tiefenbacher}},
  \bibinfo{author}{\bibfnamefont{T.}~\bibnamefont{Schmitt-Manderbach}},
  \bibinfo{author}{\bibfnamefont{H.}~\bibnamefont{Weier}},
  \bibinfo{author}{\bibfnamefont{T.}~\bibnamefont{Scheidl}},
  \bibinfo{author}{\bibfnamefont{M.}~\bibnamefont{Lindenthal}},
  \bibinfo{author}{\bibfnamefont{B.}~\bibnamefont{Blauen\-steiner}},
  \bibinfo{author}{\bibfnamefont{T.}~\bibnamefont{Jennewein}},
  \bibinfo{author}{\bibfnamefont{J.}~\bibnamefont{Perdigues}},
  \bibinfo{author}{\bibfnamefont{P.}~\bibnamefont{Trojek}},
  \bibnamefont{et~al.}, \bibinfo{journal}{Nature Physics}
  \textbf{\bibinfo{volume}{3}}, \bibinfo{pages}{481} (\bibinfo{year}{2007}).

\bibitem[{\citenamefont{Waks et~al.}(2002)\citenamefont{Waks, Zeevi, and
  Yamamoto}}]{Waks2002a}
\bibinfo{author}{\bibfnamefont{E.}~\bibnamefont{Waks}},
  \bibinfo{author}{\bibfnamefont{A.}~\bibnamefont{Zeevi}}, \bibnamefont{and}
  \bibinfo{author}{\bibfnamefont{Y.}~\bibnamefont{Yamamoto}},
  \bibinfo{journal}{Phys. Rev.~A} \textbf{\bibinfo{volume}{65}},
  \bibinfo{pages}{052310} (\bibinfo{year}{2002}).

\bibitem[{\citenamefont{Briegel et~al.}(1998)\citenamefont{Briegel, D{\"u}r,
  Cirac, and Zoller}}]{Briegel1998a}
\bibinfo{author}{\bibfnamefont{H.-J.} \bibnamefont{Briegel}},
  \bibinfo{author}{\bibfnamefont{W.}~\bibnamefont{D{\"u}r}},
  \bibinfo{author}{\bibfnamefont{J.~I.} \bibnamefont{Cirac}}, \bibnamefont{and}
  \bibinfo{author}{\bibfnamefont{P.}~\bibnamefont{Zoller}},
  \bibinfo{journal}{Phys. Rev. Lett.} \textbf{\bibinfo{volume}{81}},
  \bibinfo{pages}{5932} (\bibinfo{year}{1998}).

\bibitem[{\citenamefont{D{\"u}r et~al.}(1999)\citenamefont{D{\"u}r, Briegel,
  Cirac, and Zoller}}]{Duer1999a}
\bibinfo{author}{\bibfnamefont{W.}~\bibnamefont{D{\"u}r}},
  \bibinfo{author}{\bibfnamefont{H.-J.} \bibnamefont{Briegel}},
  \bibinfo{author}{\bibfnamefont{J.~I.} \bibnamefont{Cirac}}, \bibnamefont{and}
  \bibinfo{author}{\bibfnamefont{P.}~\bibnamefont{Zoller}},
  \bibinfo{journal}{Phys. Rev.~A} \textbf{\bibinfo{volume}{59}},
  \bibinfo{pages}{169} (\bibinfo{year}{1999}).

\bibitem[{\citenamefont{Duan et~al.}(2001)\citenamefont{Duan, Lukin, Cirac, and
  Zoller}}]{Duan2001a}
\bibinfo{author}{\bibfnamefont{L.-M.} \bibnamefont{Duan}},
  \bibinfo{author}{\bibfnamefont{M.~D.} \bibnamefont{Lukin}},
  \bibinfo{author}{\bibfnamefont{J.~I.} \bibnamefont{Cirac}}, \bibnamefont{and}
  \bibinfo{author}{\bibfnamefont{P.}~\bibnamefont{Zoller}},
  \bibinfo{journal}{Nature} \textbf{\bibinfo{volume}{414}},
  \bibinfo{pages}{413} (\bibinfo{year}{2001}).

\bibitem[{\citenamefont{Sangouard et~al.}(2008)\citenamefont{Sangouard, Simon,
  Zhao, Chen, de~Riedmatten, Pan, and Gisin}}]{Sangouard2008a}
\bibinfo{author}{\bibfnamefont{N.}~\bibnamefont{Sangouard}},
  \bibinfo{author}{\bibfnamefont{C.}~\bibnamefont{Simon}},
  \bibinfo{author}{\bibfnamefont{B.}~\bibnamefont{Zhao}},
  \bibinfo{author}{\bibfnamefont{Y.-A.} \bibnamefont{Chen}},
  \bibinfo{author}{\bibfnamefont{H.}~\bibnamefont{de~Riedmatten}},
  \bibinfo{author}{\bibfnamefont{J.-W.} \bibnamefont{Pan}}, \bibnamefont{and}
  \bibinfo{author}{\bibfnamefont{N.}~\bibnamefont{Gisin}},
  \bibinfo{journal}{Phys. Rev.~A} \textbf{\bibinfo{volume}{77}},
  \bibinfo{pages}{062301} (\bibinfo{year}{2008}).

\bibitem[{\citenamefont{Bennett et~al.}(1996)\citenamefont{Bennett, Brassard,
  Popescu, Schumacher, Smolin, and Wootters}}]{Bennett1996a}
\bibinfo{author}{\bibfnamefont{C.~H.} \bibnamefont{Bennett}},
  \bibinfo{author}{\bibfnamefont{G.}~\bibnamefont{Brassard}},
  \bibinfo{author}{\bibfnamefont{S.}~\bibnamefont{Popescu}},
  \bibinfo{author}{\bibfnamefont{B.}~\bibnamefont{Schumacher}},
  \bibinfo{author}{\bibfnamefont{J.~A.} \bibnamefont{Smolin}},
  \bibnamefont{and} \bibinfo{author}{\bibfnamefont{W.~K.}
  \bibnamefont{Wootters}}, \bibinfo{journal}{Phys. Rev. Lett.}
  \textbf{\bibinfo{volume}{76}}, \bibinfo{pages}{722} (\bibinfo{year}{1996}).

\bibitem[{\citenamefont{Deutsch et~al.}(1996)\citenamefont{Deutsch, Ekert,
  Jozsa, Macchiavello, Popescu, and Sanpera}}]{Deutsch1996a}
\bibinfo{author}{\bibfnamefont{D.}~\bibnamefont{Deutsch}},
  \bibinfo{author}{\bibfnamefont{A.}~\bibnamefont{Ekert}},
  \bibinfo{author}{\bibfnamefont{R.}~\bibnamefont{Jozsa}},
  \bibinfo{author}{\bibfnamefont{C.}~\bibnamefont{Macchiavello}},
  \bibinfo{author}{\bibfnamefont{S.}~\bibnamefont{Popescu}}, \bibnamefont{and}
  \bibinfo{author}{\bibfnamefont{A.}~\bibnamefont{Sanpera}},
  \bibinfo{journal}{Phys. Rev. Lett.} \textbf{\bibinfo{volume}{77}},
  \bibinfo{pages}{2818} (\bibinfo{year}{1996}).

\bibitem[{\citenamefont{{\.Z}ukowski et~al.}(1993)\citenamefont{{\.Z}ukowski,
  Zeilinger, Horne, and Ekert}}]{Zukowski1993a}
\bibinfo{author}{\bibfnamefont{M.}~\bibnamefont{{\.Z}ukowski}},
  \bibinfo{author}{\bibfnamefont{A.}~\bibnamefont{Zeilinger}},
  \bibinfo{author}{\bibfnamefont{M.~A.} \bibnamefont{Horne}}, \bibnamefont{and}
  \bibinfo{author}{\bibfnamefont{A.~K.} \bibnamefont{Ekert}},
  \bibinfo{journal}{Phys. Rev. Lett.} \textbf{\bibinfo{volume}{71}},
  \bibinfo{pages}{4287} (\bibinfo{year}{1993}).

\bibitem[{\citenamefont{Pan et~al.}(1998)\citenamefont{Pan, Bouwmeester,
  Wein\-furter, and Zeilinger}}]{Pan1998a}
\bibinfo{author}{\bibfnamefont{J.-W.} \bibnamefont{Pan}},
  \bibinfo{author}{\bibfnamefont{D.}~\bibnamefont{Bouwmeester}},
  \bibinfo{author}{\bibfnamefont{H.}~\bibnamefont{Wein\-furter}},
  \bibnamefont{and}
  \bibinfo{author}{\bibfnamefont{A.}~\bibnamefont{Zeilinger}},
  \bibinfo{journal}{Phys. Rev. Lett.} \textbf{\bibinfo{volume}{80}},
  \bibinfo{pages}{3891} (\bibinfo{year}{1998}).

\bibitem[{\citenamefont{Jennewein et~al.}(2001)\citenamefont{Jennewein, Weihs,
  Pan, and Zeilinger}}]{Jennewein2001a}
\bibinfo{author}{\bibfnamefont{T.}~\bibnamefont{Jennewein}},
  \bibinfo{author}{\bibfnamefont{G.}~\bibnamefont{Weihs}},
  \bibinfo{author}{\bibfnamefont{J.-W.} \bibnamefont{Pan}}, \bibnamefont{and}
  \bibinfo{author}{\bibfnamefont{A.}~\bibnamefont{Zeilinger}},
  \bibinfo{journal}{Phys. Rev. Lett.} \textbf{\bibinfo{volume}{88}},
  \bibinfo{pages}{017903} (\bibinfo{year}{2001}).

\bibitem[{\citenamefont{Pan et~al.}(2003)\citenamefont{Pan, Gasparoni, Ursin,
  Weihs, and Zeilinger}}]{Pan2003a}
\bibinfo{author}{\bibfnamefont{J.-W.} \bibnamefont{Pan}},
  \bibinfo{author}{\bibfnamefont{S.}~\bibnamefont{Gasparoni}},
  \bibinfo{author}{\bibfnamefont{R.}~\bibnamefont{Ursin}},
  \bibinfo{author}{\bibfnamefont{G.}~\bibnamefont{Weihs}}, \bibnamefont{and}
  \bibinfo{author}{\bibfnamefont{A.}~\bibnamefont{Zeilinger}},
  \bibinfo{journal}{Nature} \textbf{\bibinfo{volume}{423}},
  \bibinfo{pages}{417} (\bibinfo{year}{2003}).

\bibitem[{\citenamefont{Matsukevich and Kuzmich}(2004)}]{Matsukevich2004a}
\bibinfo{author}{\bibfnamefont{D.~N.} \bibnamefont{Matsukevich}}
  \bibnamefont{and} \bibinfo{author}{\bibfnamefont{A.}~\bibnamefont{Kuzmich}},
  \bibinfo{journal}{Science} \textbf{\bibinfo{volume}{306}},
  \bibinfo{pages}{663} (\bibinfo{year}{2004}).

\bibitem[{\citenamefont{Jennewein et~al.}(2004)\citenamefont{Jennewein, Ursin,
  Aspelmeyer, and Zeilinger}}]{Jennewein2004a}
\bibinfo{author}{\bibfnamefont{T.}~\bibnamefont{Jennewein}},
  \bibinfo{author}{\bibfnamefont{R.}~\bibnamefont{Ursin}},
  \bibinfo{author}{\bibfnamefont{M.}~\bibnamefont{Aspelmeyer}},
  \bibnamefont{and} \bibinfo{author}{\bibfnamefont{A.}~\bibnamefont{Zeilinger}}
  (\bibinfo{year}{2004}), \bibinfo{note}{{q}uant-ph/0409008}.

\bibitem[{\citenamefont{Walther et~al.}(2005)\citenamefont{Walther, Resch,
  Brukner, Steinberg, Pan, and Zeilinger}}]{Walther2005a}
\bibinfo{author}{\bibfnamefont{P.}~\bibnamefont{Walther}},
  \bibinfo{author}{\bibfnamefont{K.~J.} \bibnamefont{Resch}},
  \bibinfo{author}{\bibfnamefont{{\v C}.}~\bibnamefont{Brukner}},
  \bibinfo{author}{\bibfnamefont{A.~M.} \bibnamefont{Steinberg}},
  \bibinfo{author}{\bibfnamefont{J.-W.} \bibnamefont{Pan}}, \bibnamefont{and}
  \bibinfo{author}{\bibfnamefont{A.}~\bibnamefont{Zeilinger}},
  \bibinfo{journal}{Phys. Rev. Lett.} \textbf{\bibinfo{volume}{94}},
  \bibinfo{pages}{040504} (\bibinfo{year}{2005}).

\bibitem[{\citenamefont{Yuan et~al.}(2008)\citenamefont{Yuan, Chen, Zhao, Chen,
  Schmiedmayer, and Pan}}]{Yuan2008a}
\bibinfo{author}{\bibfnamefont{Z.-S.} \bibnamefont{Yuan}},
  \bibinfo{author}{\bibfnamefont{Y.-A.} \bibnamefont{Chen}},
  \bibinfo{author}{\bibfnamefont{B.}~\bibnamefont{Zhao}},
  \bibinfo{author}{\bibfnamefont{S.}~\bibnamefont{Chen}},
  \bibinfo{author}{\bibfnamefont{J.}~\bibnamefont{Schmiedmayer}},
  \bibnamefont{and} \bibinfo{author}{\bibfnamefont{J.-W.} \bibnamefont{Pan}},
  \bibinfo{journal}{Nature} \textbf{\bibinfo{volume}{454}},
  \bibinfo{pages}{1098} (\bibinfo{year}{2008}).

\bibitem[{\citenamefont{Choi et~al.}(2008)\citenamefont{Choi, Deng, Laurat, and
  Kimble}}]{Choi2008a}
\bibinfo{author}{\bibfnamefont{K.~S.} \bibnamefont{Choi}},
  \bibinfo{author}{\bibfnamefont{H.}~\bibnamefont{Deng}},
  \bibinfo{author}{\bibfnamefont{J.}~\bibnamefont{Laurat}}, \bibnamefont{and}
  \bibinfo{author}{\bibfnamefont{H.~J.} \bibnamefont{Kimble}},
  \bibinfo{journal}{Nature} \textbf{\bibinfo{volume}{452}}, \bibinfo{pages}{67}
  (\bibinfo{year}{2008}).

\bibitem[{\citenamefont{Matsukevich et~al.}(2008)\citenamefont{Matsukevich,
  Maunz, Moehring, Olmschenk, and Monroe}}]{Matsukevich2008a}
\bibinfo{author}{\bibfnamefont{D.~N.} \bibnamefont{Matsukevich}},
  \bibinfo{author}{\bibfnamefont{P.}~\bibnamefont{Maunz}},
  \bibinfo{author}{\bibfnamefont{D.~L.} \bibnamefont{Moehring}},
  \bibinfo{author}{\bibfnamefont{S.}~\bibnamefont{Olmschenk}},
  \bibnamefont{and} \bibinfo{author}{\bibfnamefont{C.}~\bibnamefont{Monroe}},
  \bibinfo{journal}{Phys. Rev. Lett.} \textbf{\bibinfo{volume}{100}},
  \bibinfo{pages}{150404} (\bibinfo{year}{2008}).

\bibitem[{\citenamefont{Greenberger
  et~al.}(2008{\natexlab{a}})\citenamefont{Greenberger, Horne, Zeilinger, and
  {\.Z}ukowski}}]{Greenberger2008b}
\bibinfo{author}{\bibfnamefont{D.~M.} \bibnamefont{Greenberger}},
  \bibinfo{author}{\bibfnamefont{M.}~\bibnamefont{Horne}},
  \bibinfo{author}{\bibfnamefont{A.}~\bibnamefont{Zeilinger}},
  \bibnamefont{and}
  \bibinfo{author}{\bibfnamefont{M.}~\bibnamefont{{\.Z}ukowski}},
  \bibinfo{journal}{Phys. Rev.~A} \textbf{\bibinfo{volume}{78}},
  \bibinfo{pages}{022111} (\bibinfo{year}{2008}{\natexlab{a}}).

\bibitem[{\citenamefont{Bell}(1964)}]{Bell1964a}
\bibinfo{author}{\bibfnamefont{J.~S.} \bibnamefont{Bell}},
  \bibinfo{journal}{Physics} \textbf{\bibinfo{volume}{1}}, \bibinfo{pages}{195}
  (\bibinfo{year}{1964}).

\bibitem[{\citenamefont{Aspect et~al.}(1982)\citenamefont{Aspect, Dalibard, and
  Roger}}]{Aspect1982b}
\bibinfo{author}{\bibfnamefont{A.}~\bibnamefont{Aspect}},
  \bibinfo{author}{\bibfnamefont{J.}~\bibnamefont{Dalibard}}, \bibnamefont{and}
  \bibinfo{author}{\bibfnamefont{G.}~\bibnamefont{Roger}},
  \bibinfo{journal}{Phys. Rev. Lett.} \textbf{\bibinfo{volume}{49}},
  \bibinfo{pages}{1804} (\bibinfo{year}{1982}).

\bibitem[{\citenamefont{Weihs et~al.}(1998)\citenamefont{Weihs, Jennewein,
  Simon, Weinfurter, and Zeilinger}}]{Weihs1998a}
\bibinfo{author}{\bibfnamefont{G.}~\bibnamefont{Weihs}},
  \bibinfo{author}{\bibfnamefont{T.}~\bibnamefont{Jennewein}},
  \bibinfo{author}{\bibfnamefont{C.}~\bibnamefont{Simon}},
  \bibinfo{author}{\bibfnamefont{H.}~\bibnamefont{Weinfurter}},
  \bibnamefont{and}
  \bibinfo{author}{\bibfnamefont{A.}~\bibnamefont{Zeilinger}},
  \bibinfo{journal}{Phys. Rev. Lett.} \textbf{\bibinfo{volume}{81}},
  \bibinfo{pages}{5039} (\bibinfo{year}{1998}).

\bibitem[{\citenamefont{Rowe et~al.}(2001)\citenamefont{Rowe, Kielpinski,
  Meyer, Sackett, Itano, Monroe, and Wineland}}]{Rowe2001a}
\bibinfo{author}{\bibfnamefont{M.~A.} \bibnamefont{Rowe}},
  \bibinfo{author}{\bibfnamefont{D.}~\bibnamefont{Kielpinski}},
  \bibinfo{author}{\bibfnamefont{V.}~\bibnamefont{Meyer}},
  \bibinfo{author}{\bibfnamefont{C.~A.} \bibnamefont{Sackett}},
  \bibinfo{author}{\bibfnamefont{W.~M.} \bibnamefont{Itano}},
  \bibinfo{author}{\bibfnamefont{C.}~\bibnamefont{Monroe}}, \bibnamefont{and}
  \bibinfo{author}{\bibfnamefont{D.~J.} \bibnamefont{Wineland}},
  \bibinfo{journal}{Nature} \textbf{\bibinfo{volume}{409}},
  \bibinfo{pages}{791} (\bibinfo{year}{2001}).

\bibitem[{\citenamefont{Yang et~al.}(2006)\citenamefont{Yang, Zhang, Chen, Lu,
  Yin, Pan, Wei, Tian, and Zhang}}]{Yang2006a}
\bibinfo{author}{\bibfnamefont{T.}~\bibnamefont{Yang}},
  \bibinfo{author}{\bibfnamefont{Q.}~\bibnamefont{Zhang}},
  \bibinfo{author}{\bibfnamefont{T.-Y.} \bibnamefont{Chen}},
  \bibinfo{author}{\bibfnamefont{S.}~\bibnamefont{Lu}},
  \bibinfo{author}{\bibfnamefont{J.}~\bibnamefont{Yin}},
  \bibinfo{author}{\bibfnamefont{J.-W.} \bibnamefont{Pan}},
  \bibinfo{author}{\bibfnamefont{Z.-Y.} \bibnamefont{Wei}},
  \bibinfo{author}{\bibfnamefont{J.-R.} \bibnamefont{Tian}}, \bibnamefont{and}
  \bibinfo{author}{\bibfnamefont{J.}~\bibnamefont{Zhang}},
  \bibinfo{journal}{Phys. Rev. Lett.} \textbf{\bibinfo{volume}{96}},
  \bibinfo{pages}{110501} (\bibinfo{year}{2006}).

\bibitem[{\citenamefont{Kaltenbaek et~al.}(2006)\citenamefont{Kaltenbaek,
  Blauensteiner, {\.Z}ukowski, Aspelmeyer, and Zeilinger}}]{Kaltenbaek2006a}
\bibinfo{author}{\bibfnamefont{R.}~\bibnamefont{Kaltenbaek}},
  \bibinfo{author}{\bibfnamefont{B.}~\bibnamefont{Blauensteiner}},
  \bibinfo{author}{\bibfnamefont{M.}~\bibnamefont{{\.Z}ukowski}},
  \bibinfo{author}{\bibfnamefont{M.}~\bibnamefont{Aspelmeyer}},
  \bibnamefont{and}
  \bibinfo{author}{\bibfnamefont{A.}~\bibnamefont{Zeilinger}},
  \bibinfo{journal}{Phys. Rev. Lett.} \textbf{\bibinfo{volume}{96}},
  \bibinfo{pages}{240502} (\bibinfo{year}{2006}).

\bibitem[{\citenamefont{Halder et~al.}(2007)\citenamefont{Halder, Beveratos,
  Gisin, Scarani, Simon, and Zbinden}}]{Halder2007a}
\bibinfo{author}{\bibfnamefont{M.}~\bibnamefont{Halder}},
  \bibinfo{author}{\bibfnamefont{A.}~\bibnamefont{Beveratos}},
  \bibinfo{author}{\bibfnamefont{N.}~\bibnamefont{Gisin}},
  \bibinfo{author}{\bibfnamefont{V.}~\bibnamefont{Scarani}},
  \bibinfo{author}{\bibfnamefont{C.}~\bibnamefont{Simon}}, \bibnamefont{and}
  \bibinfo{author}{\bibfnamefont{H.}~\bibnamefont{Zbinden}},
  \bibinfo{journal}{Nature Physics} \textbf{\bibinfo{volume}{3}},
  \bibinfo{pages}{692} (\bibinfo{year}{2007}).

\bibitem[{\citenamefont{Bennett et~al.}(1993)\citenamefont{Bennett, Brassard,
  Cr{\' e}peau, Jozsa, Peres, and Wootters}}]{Bennett1993a}
\bibinfo{author}{\bibfnamefont{C.~H.} \bibnamefont{Bennett}},
  \bibinfo{author}{\bibfnamefont{G.}~\bibnamefont{Brassard}},
  \bibinfo{author}{\bibfnamefont{C.}~\bibnamefont{Cr{\' e}peau}},
  \bibinfo{author}{\bibfnamefont{R.}~\bibnamefont{Jozsa}},
  \bibinfo{author}{\bibfnamefont{A.}~\bibnamefont{Peres}}, \bibnamefont{and}
  \bibinfo{author}{\bibfnamefont{W.~K.} \bibnamefont{Wootters}},
  \bibinfo{journal}{Phys. Rev. Lett.} \textbf{\bibinfo{volume}{70}},
  \bibinfo{pages}{1895} (\bibinfo{year}{1993}).

\bibitem[{\citenamefont{Ekert}(1991)}]{Ekert1991a}
\bibinfo{author}{\bibfnamefont{A.~K.} \bibnamefont{Ekert}},
  \bibinfo{journal}{Phys. Rev. Lett.} \textbf{\bibinfo{volume}{67}},
  \bibinfo{pages}{661} (\bibinfo{year}{1991}).

\bibitem[{\citenamefont{Bennett et~al.}(1992)\citenamefont{Bennett, Brassard,
  and Mermin}}]{Bennett1992b}
\bibinfo{author}{\bibfnamefont{C.~H.} \bibnamefont{Bennett}},
  \bibinfo{author}{\bibfnamefont{G.}~\bibnamefont{Brassard}}, \bibnamefont{and}
  \bibinfo{author}{\bibfnamefont{N.~D.} \bibnamefont{Mermin}},
  \bibinfo{journal}{Phys. Rev. Lett.} \textbf{\bibinfo{volume}{68}},
  \bibinfo{pages}{557} (\bibinfo{year}{1992}).

\bibitem[{\citenamefont{Clauser et~al.}(1969)\citenamefont{Clauser, Horne,
  Shimony, and Holt}}]{Clauser1969a}
\bibinfo{author}{\bibfnamefont{J.~F.} \bibnamefont{Clauser}},
  \bibinfo{author}{\bibfnamefont{M.~A.} \bibnamefont{Horne}},
  \bibinfo{author}{\bibfnamefont{A.}~\bibnamefont{Shimony}}, \bibnamefont{and}
  \bibinfo{author}{\bibfnamefont{R.~A.} \bibnamefont{Holt}},
  \bibinfo{journal}{Phys. Rev. Lett.} \textbf{\bibinfo{volume}{23}},
  \bibinfo{pages}{880} (\bibinfo{year}{1969}).

\bibitem[{\citenamefont{{\. Z}ukowski et~al.}(1995)\citenamefont{{\. Z}ukowski,
  Zeilinger, and Weinfurter}}]{Zukowski1995a}
\bibinfo{author}{\bibfnamefont{M.}~\bibnamefont{{\. Z}ukowski}},
  \bibinfo{author}{\bibfnamefont{A.}~\bibnamefont{Zeilinger}},
  \bibnamefont{and}
  \bibinfo{author}{\bibfnamefont{H.}~\bibnamefont{Weinfurter}},
  \bibinfo{journal}{Ann. N. Y. Acad. Sci.} \textbf{\bibinfo{volume}{755}},
  \bibinfo{pages}{91} (\bibinfo{year}{1995}).

\bibitem[{\citenamefont{Hong et~al.}(1987)\citenamefont{Hong, Ou, and
  Mandel}}]{Hong1987a}
\bibinfo{author}{\bibfnamefont{C.~K.} \bibnamefont{Hong}},
  \bibinfo{author}{\bibfnamefont{Z.~Y.} \bibnamefont{Ou}}, \bibnamefont{and}
  \bibinfo{author}{\bibfnamefont{L.}~\bibnamefont{Mandel}},
  \bibinfo{journal}{Phys. Rev. Lett.} \textbf{\bibinfo{volume}{59}},
  \bibinfo{pages}{2044} (\bibinfo{year}{1987}).

\bibitem[{\citenamefont{Braunstein and Mann}(1995)}]{Braunstein1995a}
\bibinfo{author}{\bibfnamefont{S.~L.} \bibnamefont{Braunstein}}
  \bibnamefont{and} \bibinfo{author}{\bibfnamefont{A.}~\bibnamefont{Mann}},
  \bibinfo{journal}{Phys. Rev.~A} \textbf{\bibinfo{volume}{51}},
  \bibinfo{pages}{R1727} (\bibinfo{year}{1995}).

\bibitem[{\citenamefont{Mattle et~al.}(1996)\citenamefont{Mattle, Weinfurter,
  Kwiat, and Zeilinger}}]{Mattle1996a}
\bibinfo{author}{\bibfnamefont{K.}~\bibnamefont{Mattle}},
  \bibinfo{author}{\bibfnamefont{H.}~\bibnamefont{Weinfurter}},
  \bibinfo{author}{\bibfnamefont{P.~G.} \bibnamefont{Kwiat}}, \bibnamefont{and}
  \bibinfo{author}{\bibfnamefont{A.}~\bibnamefont{Zeilinger}},
  \bibinfo{journal}{Phys. Rev. Lett.} \textbf{\bibinfo{volume}{76}},
  \bibinfo{pages}{4656} (\bibinfo{year}{1996}).

\bibitem[{\citenamefont{Calsamiglia and L{\"
  u}tkenhaus}(2001)}]{Calsamiglia2001a}
\bibinfo{author}{\bibfnamefont{J.}~\bibnamefont{Calsamiglia}} \bibnamefont{and}
  \bibinfo{author}{\bibfnamefont{N.}~\bibnamefont{L{\" u}tkenhaus}},
  \bibinfo{journal}{Appl. Phys.~B} \textbf{\bibinfo{volume}{72}},
  \bibinfo{pages}{67} (\bibinfo{year}{2001}).

\bibitem[{\citenamefont{Horodecki et~al.}(1997)\citenamefont{Horodecki,
  Horodecki, and Horodecki}}]{Horodecki1997a}
\bibinfo{author}{\bibfnamefont{M.}~\bibnamefont{Horodecki}},
  \bibinfo{author}{\bibfnamefont{P.}~\bibnamefont{Horodecki}},
  \bibnamefont{and}
  \bibinfo{author}{\bibfnamefont{R.}~\bibnamefont{Horodecki}},
  \bibinfo{journal}{Phys. Rev. Lett.} \textbf{\bibinfo{volume}{78}},
  \bibinfo{pages}{574} (\bibinfo{year}{1997}).

\bibitem[{\citenamefont{Werner}(1989)}]{Werner1989a}
\bibinfo{author}{\bibfnamefont{R.~F.} \bibnamefont{Werner}},
  \bibinfo{journal}{Phys. Rev.~A} \textbf{\bibinfo{volume}{40}},
  \bibinfo{pages}{4277} (\bibinfo{year}{1989}).

\bibitem[{\citenamefont{Yurke and Stoler}(1992)}]{Yurke1992b}
\bibinfo{author}{\bibfnamefont{B.}~\bibnamefont{Yurke}} \bibnamefont{and}
  \bibinfo{author}{\bibfnamefont{D.}~\bibnamefont{Stoler}},
  \bibinfo{journal}{Phys. Rev.~A} \textbf{\bibinfo{volume}{46}},
  \bibinfo{pages}{2229} (\bibinfo{year}{1992}).

\bibitem[{\citenamefont{Greenberger
  et~al.}(2008{\natexlab{b}})\citenamefont{Greenberger, Horne, and
  Zeilinger}}]{Greenberger2008a}
\bibinfo{author}{\bibfnamefont{D.~M.} \bibnamefont{Greenberger}},
  \bibinfo{author}{\bibfnamefont{M.}~\bibnamefont{Horne}}, \bibnamefont{and}
  \bibinfo{author}{\bibfnamefont{A.}~\bibnamefont{Zeilinger}},
  \bibinfo{journal}{Phys. Rev.~A} \textbf{\bibinfo{volume}{78}},
  \bibinfo{pages}{022110} (\bibinfo{year}{2008}{\natexlab{b}}).

\bibitem[{\citenamefont{Lutes}(1987)}]{Lutes1987a}
\bibinfo{author}{\bibfnamefont{G.}~\bibnamefont{Lutes}}, in
  \emph{\bibinfo{booktitle}{{P}roceedings of the 41{$^\textrm{st}$} {A}nnual
  {S}ymposium\\ on {F}requency {C}ontrol}} (\bibinfo{year}{1987}), pp.
  \bibinfo{pages}{161--166}.

\bibitem[{\citenamefont{Celano et~al.}(2002)\citenamefont{Celano, Stein,
  Gifford, Mesander, and Ramsey}}]{Celano2002a}
\bibinfo{author}{\bibfnamefont{T.~P.} \bibnamefont{Celano}},
  \bibinfo{author}{\bibfnamefont{S.~R.} \bibnamefont{Stein}},
  \bibinfo{author}{\bibfnamefont{G.~A.} \bibnamefont{Gifford}},
  \bibinfo{author}{\bibfnamefont{B.~A.} \bibnamefont{Mesander}},
  \bibnamefont{and} \bibinfo{author}{\bibfnamefont{B.~J.}
  \bibnamefont{Ramsey}}, in \emph{\bibinfo{booktitle}{{F}requency {C}ontrol
  {S}ymposium and\\ {PDA} {E}xhibition}} (\bibinfo{organization}{IEEE
  International}, \bibinfo{year}{2002}), pp. \bibinfo{pages}{510--516}.

\bibitem[{\citenamefont{Hudson et~al.}(2006)\citenamefont{Hudson, Foreman,
  Cundiff, and Ye}}]{Hudson2006a}
\bibinfo{author}{\bibfnamefont{D.~D.} \bibnamefont{Hudson}},
  \bibinfo{author}{\bibfnamefont{S.~M.} \bibnamefont{Foreman}},
  \bibinfo{author}{\bibfnamefont{S.~T.} \bibnamefont{Cundiff}},
  \bibnamefont{and} \bibinfo{author}{\bibfnamefont{J.}~\bibnamefont{Ye}},
  \bibinfo{journal}{Opt. Lett.} \textbf{\bibinfo{volume}{31}},
  \bibinfo{pages}{1951} (\bibinfo{year}{2006}).

\bibitem[{\citenamefont{Kim et~al.}(2008)\citenamefont{Kim, Cox, Chen, and
  K{\"a}rtner}}]{Kim2008a}
\bibinfo{author}{\bibfnamefont{J.}~\bibnamefont{Kim}},
  \bibinfo{author}{\bibfnamefont{J.~A.} \bibnamefont{Cox}},
  \bibinfo{author}{\bibfnamefont{J.}~\bibnamefont{Chen}}, \bibnamefont{and}
  \bibinfo{author}{\bibfnamefont{F.~X.} \bibnamefont{K{\"a}rtner}}
  (\bibinfo{year}{2008}), \bibinfo{note}{{N}ature {P}hotonics, {A}dvance
  {O}nline {P}ubication, 2. {N}ovember, 2008, doi: 10.1038/nphoton.2008.225}.

\bibitem[{\citenamefont{Fedrizzi et~al.}(2007)\citenamefont{Fedrizzi, Herbst,
  Poppe, Jennewein, and Zeilinger}}]{Fedrizzi2007a}
\bibinfo{author}{\bibfnamefont{A.}~\bibnamefont{Fedrizzi}},
  \bibinfo{author}{\bibfnamefont{T.}~\bibnamefont{Herbst}},
  \bibinfo{author}{\bibfnamefont{A.}~\bibnamefont{Poppe}},
  \bibinfo{author}{\bibfnamefont{T.}~\bibnamefont{Jennewein}},
  \bibnamefont{and}
  \bibinfo{author}{\bibfnamefont{A.}~\bibnamefont{Zeilinger}},
  \bibinfo{journal}{Opt. Exp.} \textbf{\bibinfo{volume}{15}},
  \bibinfo{pages}{15377} (\bibinfo{year}{2007}).

\bibitem[{\citenamefont{Bao et~al.}(2008)\citenamefont{Bao, Qian, Yang, Zhang,
  Chen, Yang, and Pan}}]{Bao2008a}
\bibinfo{author}{\bibfnamefont{X.-H.} \bibnamefont{Bao}},
  \bibinfo{author}{\bibfnamefont{Y.}~\bibnamefont{Qian}},
  \bibinfo{author}{\bibfnamefont{J.}~\bibnamefont{Yang}},
  \bibinfo{author}{\bibfnamefont{H.}~\bibnamefont{Zhang}},
  \bibinfo{author}{\bibfnamefont{Z.-B.} \bibnamefont{Chen}},
  \bibinfo{author}{\bibfnamefont{T.}~\bibnamefont{Yang}}, \bibnamefont{and}
  \bibinfo{author}{\bibfnamefont{J.-W.} \bibnamefont{Pan}},
  \bibinfo{journal}{Phys. Rev. Lett.} \textbf{\bibinfo{volume}{101}},
  \bibinfo{pages}{190501} (\bibinfo{year}{2008}).

\end{thebibliography}

\end{document}